\documentclass[slac_one]{revtex4}
\usepackage{graphicx}
\usepackage{fancyhdr}
\usepackage{relsize}
\def\babar{\mbox{\slshape B\kern-0.1em{\smaller A}\kern-0.1em
    B\kern-0.1em{\smaller A\kern-0.2em R}}}
\def\CP                {\ensuremath{C\!P}}

\begin{document}

\title{Charmless Three-body $B$ Decays at \babar} 

%

\author{F. Liu} 
\affiliation{UC Riverside at SLAC, 2575 Sand Hill Road,
 Menlo Park, CA 94025, USA\\ 
Representing the \babar\ Collaboration}

\begin{abstract}
We present Dalitz-plot analyses of $B^+\to K^+\pi^-\pi^+$ and 
$B^0\to K^+\pi^-\pi^0$ using the data  sample collected at the 
$\Upsilon(4S)$ by the \babar\ detector. We have found evidence for
direct $CP$-violation in the decay 
$B^\pm\to\rho^0 K^\pm$, with a $CP$-violation parameter 
${\cal A}_{CP}={\Gamma_{B^+}-\Gamma_{B^-}\over \Gamma_{B^+}+\Gamma_{B^-}}=
\left(+44\pm10\pm4^{+5}_{-13}\right)\%$, where $\Gamma_{B^\pm}$ 
are the decay rates. The uncertainties are statistical, systematic, 
and model-dependent, respectively.
We also search for the suppressed decays $B^+\to K^-\pi^+\pi^+$ and 
$K^+K^+\pi^-$ and improve upper limits on the decay branching fractions. 
\end{abstract}

\maketitle

\thispagestyle{fancy}

\section{INTRODUCTION} 
In the Standard Model (SM), 
violation of $CP$ symmetry is a consequence of
the complex phase of the Cabibbo-Kobayashi-Maskawa (CKM)
quark-mixing matrix~\cite{ckm}. 
Comprehensive tests of the SM
$CP$-violation mechanism require precise measurements of the
three sides and three angles of the CKM unitarity triangle.
Although $CP$-violation in the neutral $K^0$ meson~\cite{KaonCP} 
and $B^0$ meson~\cite{BCP} has been well-established,
it is known 
to be too small to account for the matter-dominated  Universe,
direct \CP-violation  would help to explain
the dominance of matter in the Universe~\cite{Nature}.
Direct \CP-asymmetry in $B^\pm\to\rho^0 K^\pm$ has been 
suggested~\cite{rhoKCP} and searched at \babar\ and BELLE~\cite{DCPBB} 
through Dalitz-plot analysis of $B^+\to K^+\pi^+\pi^-$ (charge conjugate
decay is implied throughout this paper.)
Direct \CP-asymmetries in $B\to K^*\pi$, $K_0^*\pi$, 
and $K_2^*\pi$ are expected to be small in the SM. 
Moreover, the relative weak phase between tree and penguin diagrams in 
$B\to K\pi\pi$ decays is the CKM angle 
$\gamma\equiv arg\left(-V_{ud}V_{ub}^*/V_{cd}V_{cb}^*\right)$. Therefore
a set of Dalitz-plot
analyses of $B\to K\pi\pi$ can provide a relatively clean determination 
of $\gamma$~\cite{KpipiDP}.

Compared to the penguin transitions $\bar b\to q\bar q \bar d$ and
$\bar b\to q\bar q \bar s$, the decay rates for the wrong sign 
decays $B^+\to K^-\pi^+\pi^+$ and $K^+K^+\pi^-$ 
via $\bar b \to \bar d\bar d s$
and $\bar b\to \bar s \bar s d$ transitions are further 
suppressed by $\left| V_{td}V_{ts}^*\right|^2 \simeq {\cal O}(10^{-7})$, 
resulting in branching fractions of ${\cal O}(10^{-14})$ and
${\cal O}(10^{-11})$, respectively~\cite{Huitu}. 
Observations of these
decays would be clear evidence of the $\bar b\to \bar d\bar d s$ 
and $\bar b\to \bar s \bar sd$ transitions.

\section{ANALYSIS TECHNIQUE}

A number of intermediate states can contribute to 3-body $B\to K\pi\pi$ decays. 
Their individual contributions are obtained from a maximum likelihood (ML) 
fit to the distribution of events in the Dalitz-plot formed by two
invariant masses squared $x$ and $y$ of particle pairs. 
The total amplitudes for 3-body $B$ and $\overline B$ decays are 
given~\cite{isobar1} by:
\begin{eqnarray}
A\equiv A(x,~y) &=& \sum_j c_j F_j(x,~y); ~ ~ ~ 
\overline{A}\equiv\overline{A}(x,~y)=\sum_j \overline{c}_j \overline{F}_j(x,~y).
\end{eqnarray}
Where $c_j$ is the complex coefficient for a given intermediate state
 $j$  
with all the weak phase dependence.  The normalized 
distributions $F_j$ describe the dynamics of the decay amplitudes and
are the product of an invariant mass term 
 (relativistic Breit-Wigner in general), 
two Blatt-Weisskopf barrier form factors~\cite{blatt-weisskopf},
 and an angular function.  In the case of $f_0(980)$ ($K_0^*$), 
the mass term is replaced by the Flatt\'e (LASS) lineshape~\cite{LASS}. 
The fit fraction $FF$ of a given intermediate state $j$ with a 
partial decay width $\Gamma_j$ is given by:
\begin{equation}
{\it FF}_j\equiv {\Gamma_j\over\Gamma}=
\frac
{\displaystyle\int\!\!\int{\left(\left|c_j F_j\right|^2 + \left|\overline{c}_j \overline{F}_j\right|^2\right)} dx \, dy}
{\displaystyle\int\!\!\int{\left(\left|A\right|^2 + \left|\overline{A}\right|^2\right)} dx\, dy} \,.
\label{eq:fitfraction}
\end{equation}
The sum of all contributions is not necessarily unity due to the
interference. 
The \CP-asymmetry for a given intermediate state is defined as 
\begin{eqnarray}
  {\cal A}_{\CP\!,\,j}\equiv 
  {\Gamma_{\overline B,j}-\Gamma_{B,j}\over \Gamma_{\overline B,j}+\Gamma_{B,j}}& = & 
  \frac
  {\left|\overline{c}_j\right|^2 - \left|c_j\right|^2}
  {\left|\overline{c}_j\right|^2 + \left|c_j\right|^2}, 
  \label{eq:cpasym}
\end{eqnarray}

The signal Dalitz-plot probability density function (PDF) is constructed as
\begin{eqnarray}
{\cal P}_{sig}(x,~y,q_B)={{{1+q_B\over2}\left|A\right|^2\epsilon
 +{1-q_B\over2}\left|\overline A\right|^2\overline\epsilon}\over 
\int\int\left(\left|A\right|^2\epsilon+\left|\overline A\right|^2\overline\epsilon\right) dxdy},  
\end{eqnarray}
where $q_B$ is the charge of $B$ candidate, $\epsilon\equiv\epsilon(x,y)$ 
and $\overline\epsilon=\overline\epsilon(x,y)$
are the signal reconstruction efficiencies for $B$ and $\overline B$ events. 
The Dalitz-plot PDFs for continuum and $B\overline B$ backgrounds 
are constructed through two-dimensional histograms
 from simulated continuum and $B\overline B$ samples. 

A $B$ candidate is reconstructed through the desired decay chains of
interest.
Charged kaon and pion candidates are identified by energy loss $dE/dx$ 
information measured in a five-layer silicon vertex detector and a 40-layer
drift chamber, and the Cherenkov angle and number of photons measured in 
a detector of internally reflected Cherenkov radiation. A $\pi^0$ candidate
is formed from photon pairs measured in a CsI(Tl) crystal electromagnetic
calorimeter. The \babar\ detector is described in detail elsewhere~\cite{detector}. 

A $B$ candidate is characterized by two kinematic variables: the 
energy-substituted mass
${m_{\rm ES}}\equiv\sqrt{s/4-(p_B^*)^2}$ and the energy difference
$\Delta E\equiv E_B^*-\sqrt s/2$,
where $E_B^*$ and $p_B^*$ are the center-of-mass (CM) 
energy and momentum of the $B$ candidate, respectively, $\sqrt s$ 
is the total CM energy. 
Signal events peak at the nominal $B$ mass for $m_{\rm ES}$
and at zero for $\Delta E$.

The dominant background comes from 
continuum production $e^+e^-\to q\bar q$, where $q = u,d,s,c$.  This 
background is suppressed by requirements on event-shape variables
calculated in the CM frame.  The continuum background is further suppressed
by exploring a neural network (NN) algorithm 
based on a set of kinematical variables. 

Standard extended unbinned maximum likelihood method is used to fit data. 
The likelihood function has the form 
\begin{equation}
  {\cal L}= exp{\left(-\sum_{k} n_k\right)} \prod_{i=1}^N
  \left[\sum_{k} n_k  {\cal P}_k^i\left(x,~y,~m_{\rm ES},~ \Delta E,~NN\right)\right],
\end{equation}
where $N=\sum_{k} n_k$ is the total number of events,
$n_k$ is the fit yield of component $k$ ($k=$ signal, $B\overline B$, 
and continuum).
${\cal P}^i_k$ is the PDF for event $i$ to be identified 
as component $k$.

\section{RESULTS}
\subsection{Results on $B^+\to K^+\pi^-\pi^+$}
A Dalitz-plot analysis of the decay $B^+\to K^+\pi^-\pi^+$~\cite{DPKpipi} is 
based on a data sample of 347.5 fb$^{-1}$, containing 
$(383.2\pm4.2)\times10^6$ $B\overline B$ pairs 
recorded by the \babar\ detector at the $\Upsilon(4S)$ resonance
at the Stanford Linear Accelerator Center. 
The Dalitz-plot variables are 
$x\equiv m^2_{K^+\pi^-}$ and $y\equiv m^2_{\pi^+\pi^-}$.  
A phase-space non-resonant component  
and nine intermediate states $K^{*0}\pi^+$, 
$K_0^{*0}\pi^+$, $\rho^0K^+$, $f_0(980)K^+$, $\chi_{c0}K^+$, $K_2^{*0}\pi^+$,
$\omega K^+$, $f_2(1270)K^+$, and $f_XK^+$ are included in the ML fit. 
The fit to 12,753 selected candidate events yield $4585\pm90\pm297\pm63$
signal events and the overall direct \CP-asymmetry of 
$(2.8\pm2.0\pm2.0\pm1.2)\%$, where the uncertainties are 
statistical, systematic, and model-dependent, respectively. 
 The intermediate state $\omega K^+$ with
$\omega\to\pi^+\pi^-$ has noticeable effect on the 
$\rho^0$ lineshape and is included in the fit
 although its contribution is small.  A scalar 
particle $f_X$ and $f_2(1270)$ are necessary to 
provide better fit to the data. The  
mass and width of $f_X$ are determined to be $m_{f_X}=1479\pm8$ MeV/c$^2$ and 
$\Gamma_{f_X}=80\pm10$ MeV, respectively, where the uncertainties 
are statistical only. The fit results are summarized in 
Table~\ref{tab:conclusion-results-summary}. The statistical significance
of the direct \CP-violation is evaluated from the differnce 
$-2\Delta\ln{\cal L}$ of the nagative log-likelihood of the nominal fit and 
that of a fit where \CP-violation parameters for the given  
component are set to zero, 
the number of degrees of freedom (two in this case) is taken into account. 

The total branching fraction in Table~\ref{tab:conclusion-results-summary}
 is consistent with BELLE's 
measurement~\cite{BRBELLE}. We see evidence of direct \CP-asymmetry
of ${\cal A}_{\CP} = (+44\pm10\pm4\,^{+5}_{-13})\%$ in
$B^+\to\rho^0K^+$, consistent with the previous findings~\cite{DCPBB}. 
The statistical significance of the direct 
\CP-violation effect is found to
be $3.7\sigma$ from the change in likelihood as described above. 
As experimental systematic uncertainties are much smaller than the 
statistical errors, they do not affect
this conclusion. We have cross-checked the effect of the choice of the 
Dalitz-plot models on the significance. We find that the significance 
remains above $3\sigma$ with alternative models. 
The statistical significance of direct \CP-violation in $B^+\to f_2(1270) K^+$
is also above $3\sigma$, but it suffers from large model 
uncertainties. The direct \CP-asymmetries in  $B^+\to K^{*0}\pi^+$,
$K_0^{*0}\pi^+$, and $K_2^{*0}\pi^+$ are all consistent with
the SM expectations.

\begin{table*}[htb]
\caption
{Summary of measurements of branching fractions (averaged over charge
conjugate states) and \CP\ asymmetries.
Note that these results are not corrected for secondary branching
fractions.
The first uncertainty is statistical, the second is systematic, and the
third represents the model dependence.
The final column is the statistical significance of direct \CP\ violation
(DCPV) determined as described in the text.
}
\label{tab:conclusion-results-summary}
\begin{tabular}{lcccc} \hline
Mode & Fit fraction (\%)     & ${\cal B}(B^+\to {\rm Mode}) (10^{-6})$     
     & ${\cal A}_{\CP}$ (\%) & DCPV sig.    \\ \hline
$K^+\pi^-\pi^+$ total                 &                                         & $54.4\pm1.1\pm4.5\pm0.7$                & $2.8\pm2.0\pm2.0\pm1.2$             &              \\
\hline                                                                                                                                                            
$K^{*0}\pi^+$; $K^{*0}\to K^+\pi^-$ & $13.3\pm0.7\pm0.7\,^{+0.4}_{-0.9}$      & $7.2\pm0.4\pm0.7\,^{+0.3}_{-0.5}$       & $+3.2\pm5.2\pm1.1\,^{+1.2}_{-0.7}$  &  $0.9\sigma$ \\
$K_0^{*0}\pi^+$; $K_0^{*0}\to K^+\pi^-$ & $45.0\pm1.4\pm1.2\,^{+12.9}_{-0.2}$     & $24.5\pm0.9\pm2.1\,^{+7.0}_{-1.1}$      & $+3.2\pm3.5\pm2.0\,^{+2.7}_{-1.9}$  &  $1.2\sigma$ \\
$\rho^0K^+$; $\rho^0\to\pi^+\pi^-$      & $6.54\pm0.81\pm0.58\,^{+0.69}_{-0.26}$  & $3.56\pm0.45\pm0.43\,^{+0.38}_{-0.15}$  & $+44\pm10\pm4\,^{+5}_{-13}$         &  $3.7\sigma$ \\
$f_0(980)K^+$; $f_0(980)\to\pi^+\pi^-$       & $18.9\pm0.9\pm1.7\,^{+2.8}_{-0.6}$      & $10.3\pm0.5\pm1.3\,^{+1.5}_{-0.4}$      & $-10.6\pm5.0\pm1.1\,^{+3.4}_{-1.0}$ &  $1.8\sigma$ \\
$\chi_{c0}K^+$; $\chi_{c0}\to\pi^+\pi^-$     & $1.29\pm0.19\pm0.15\,^{+0.12}_{-0.03}$  & $0.70\pm0.10\pm0.10\,^{+0.06}_{-0.02}$  & $-14\pm15\pm3\,^{+1}_{-5}$          &  $0.5\sigma$ \\
$K^+\pi^-\pi^+$ nonresonant                  & $4.5\pm0.9\pm2.4\,^{+0.6}_{-1.5}$       & $2.4\pm0.5\pm1.3\,^{+0.3}_{-0.8}$       & ---                                 &  ---         \\
$K_2^{*0}\pi^+$; $K_2^{*0}\to K^+\pi^-$      & $3.40\pm0.75\pm0.42\,^{+0.99}_{-0.13}$  & $1.85\pm0.41\pm0.28\,^{+0.54}_{-0.08}$  & $+5\pm23\pm4\,^{+18}_{-7}$          &  $0.2\sigma$ \\
$\omega K^+$; $\omega\to\pi^+\pi^-$          & $0.17\pm0.24\pm0.03\,^{+0.05}_{-0.08}$  & $0.09\pm0.13\pm0.02\,^{+0.03}_{-0.04}$  & ---                                 &  ---         \\
$f_2(1270)K^+$; $f_2(1270)\to\pi^+\pi^-$ \ \  & \ \ $0.91\pm0.27\pm0.11\,^{+0.24}_{-0.17}$ \ \ &\ \  $0.50\pm0.15\pm0.07\,^{+0.13}_{-0.09}$ \ \ & $-85\pm22\pm13\,^{+22}_{-2}$        &  $3.5\sigma$ \\
$f_XK^+$; $f_X\to\pi^+\pi^-$                & $1.33\pm0.38\pm0.86\,^{+0.04}_{-0.14}$  & $0.73\pm0.21\pm0.47\,^{+0.02}_{-0.08}$  & $+28\pm26\pm13\,^{+7}_{-5}$         &  $0.6\sigma$ \\
\hline
\end{tabular}
\end{table*}

\subsection{Results on $B^0\to K^+\pi^-\pi^0$}
A Dalitz-plot analysis of $B^0\to K^+\pi^-\pi^0$~\cite{DPKpipi0} 
is based on a data 
sample of 413 fb$^{-1}$, corresponding to $(454\pm5)\times10^6$
$B\overline B$ pairs produced at the $\Upsilon(4S)$ resonance. 
The Dalitz-plot variables are $x\equiv m^2_{K^+\pi^-}$ and 
$y\equiv m^2_{K^+\pi^0}$. 
A phase-space non-resonant component and seven intermediate 
states $\rho^-(770)K^+$, $\rho^-(1450)K^+$, $\rho^-(1700)K^+$, 
$K^{*+}\pi^-$, $K^{*0}\pi^0$, $K_0^{*+}\pi^-$, and $K_0^{*0}\pi^0$
are included in the ML fit. 
The fit to the data yields $4583\pm122$ signal events, where 
the uncertainty is statistical only. The results are preliminary. 
The decays $B^0\to\overline D^0\pi^0$ and 
$D^-K^+$ with $\overline D^0\to K^+\pi^-$ and $D^-\to\pi^+\pi^0$ are 
included in the fit as calibration modes. The fit shows that 
the decay $B^0\to K^+\pi^-\pi^0$ is dominated by 
 $K_0^{*}\pi$ and $\rho^-(770) K^+$.

Isospin symmetry relates the amplitudes of 
$B^0\to K^{*+}\pi^-$ and $K^{*0}\pi^0$
and $\overline B^0\to K^{*-}\pi^+$ and $\overline K^{*0}\pi^0$
which form two unitarity triangles.  The orientation of the 
two triangles can be determined from a time-dependent Dalitz-plot 
analysis of $B^0\to K_S^0\pi^+\pi^-$.  
A Dalitz-plot analysis of $B^0\to K^+\pi^-\pi^0$ can extract the magnitudes 
and  phases of $K^*\pi$ and 
$\overline K^*\pi$ in $B^0$ and $\overline  B^0$ decays which are essential 
to determine the CKM angle $\gamma$ and will be updated soon. 

\subsection{Results on $B^+\to K^-\pi^+\pi^+$ and $K^+K^+\pi^-$}
The preliminary results on the search for the suppressed decays 
$B^+\to K^-\pi^+\pi^+$ and $K^+K^+\pi^-$~\cite{Wrongsign} are based 
on a data sample of 426 fb$^{-1}$ which contains $(467\pm5)\times10^6$
$B\overline B$ pairs. Since the signal yields for the two modes 
are expected to be small, no Dalitz-plot involves in the ML fit. 
We apply the ML fit to the 26,478 
$B^+\to K^-\pi^+\pi^+$ and 7,822 $B^+\to K^+K^+\pi^-$ candidate events
and find no significant signal events. We set upper limits on 
the decay branching fractions to be 
${\cal B}(B^+\to K^-\pi^+\pi^+)<9.5\times10^{-7}$ and 
${\cal B}(B^+\to K^+ K^+\pi^-)<1.6\times10^{-7}$ at 90\% confidence 
level. These two upper limits  have been improved by a factor of 
2 and 8, respectively.   

\section{SUMMARY}
We have performed a Dalitz-plot analysis of $B^+\to K^+\pi^-\pi^+$. 
A scalar particle $f_X$ with $m_{f_X}=1479\pm8$ MeV/c$^2$ and
$\Gamma_{f_X}=80\pm10$ MeV and $f_2(1270)$ are necessary to fit the 
data better. We find evidence for direct $CP$-asymmetry of 
${\cal A}_{\CP} = (+44\pm10\pm4\,^{+5}_{-13})\%$ in the decay 
$B^+\to\rho^0 K^+$ with a statistical significance of $3.7\sigma$. 
The direct \CP-asymmetries in  $B^+\to K^{*0}\pi^+$,
$K_0^{*0}\pi^+$, and $K_2^{*0}\pi^+$ are consistent with zero as 
expected. 

We have improved the Dalitz-plot analysis of $B^0\to K^+\pi^-\pi^0$.
The determination of the magnitudes and phases in $B^0\to K^*\pi$ 
and $\overline B^0\to \overline K^*\pi$ 
will be updated soon which are essential for the extraction of the 
CKM angle $\gamma$. We have improved the upper limits on
the branching fractions for the SM-suppressed 
decays $B^+\to K^-\pi^+\pi^+$ and $B^+\to K^+K^+\pi^-$ 
by a factor of 2 and 8, respectively. 
\begin{acknowledgments}
I would like to thank my \babar\ collaborators for their supports and 
the organizers of ICHEP'08 for an interesting and well-organized conference. 
\end{acknowledgments}

\end{document}